\documentclass[nofootinbib,twocolumn,superscriptaddress,preprintnumbers]{revtex4-1}
\usepackage[utf8]{inputenc}
\usepackage{amsfonts}
\usepackage{mathrsfs}
\usepackage{physics}
\usepackage{footnote}
\usepackage{scrextend}
\usepackage{leftidx}
\usepackage{soul}

\usepackage[dvipsnames]{xcolor}
\definecolor{red}{rgb}{0.9, 0,0}
\definecolor{cerulean}{rgb}{0., 0.42,0.9}
\definecolor{prettygreen}{rgb}{0., 0.55,0.3}

\usepackage{epsfig}
\usepackage{graphicx}  
\usepackage{url}
\usepackage{color}
\usepackage{hyperref}
\hypersetup{
     colorlinks   = true,
     citecolor    = red,
	 linkcolor=blue
}
\usepackage{float}
\usepackage{placeins}

\newcommand{\bfx}{{\bf x}}
\newcommand{\bfq}{{\bf q}}
\newcommand{\bfp}{{\bf p}}
\newcommand{\bfr}{{\bf r}}

\newcommand{\bfk}{{\bf k}}
\newcommand{\bfG}{{\bf G}}
\newcommand{\bfv}{{\bf v}}

\newcommand{\keV}{\mathrm{keV}}

\newcommand{\bEext}{\mathbf{E}^{\rm ext}}
\newcommand{\Eext}{E^{\rm ext}}
\newcommand{\bE}{\mathbf{E}}
\newcommand{\br}{\mathbf{r}}
\newcommand{\bR}{\mathbf{R}}
\newcommand{\bG}{\mathbf{G}}
\newcommand{\bGp}{\mathbf{G}^\prime}
\newcommand{\bk}{\mathbf{k}}
\newcommand{\bq}{\mathbf{q}}

\newcommand{\om}{\omega}

\def\beq{\begin{eqnarray}}
\def\eeq{\end{eqnarray}}
\def\bea{\begin{eqnarray}}
\def\eea{\end{eqnarray}}

\definecolor{red}{rgb}{0.9, 0,0}
\definecolor{cerulean}{rgb}{0., 0.5,0.8}

\graphicspath{{./figures/}}

\begin{document}
\preprint{CERN-TH-2021-013}
\title{
Dark matter--electron scattering in dielectrics
}

\author{Simon Knapen}
\email{simon.knapen@cern.ch}
\affiliation{CERN, Theoretical Physics Department, Geneva, Switzerland}
\author{Jonathan Kozaczuk}
\email{jkozaczuk@physics.ucsd.edu}
\affiliation{Department of Physics, University of California, San Diego, CA 92093, USA}
\author{Tongyan Lin}
\email{tongyan@physics.ucsd.edu}
\affiliation{Department of Physics, University of California, San Diego, CA 92093, USA}

\date{\today}

\begin{abstract}
A number of direct detection experiments are searching for electron excitations created by scattering of sub-GeV dark matter. We present an alternate formulation of dark matter-electron scattering in terms of the dielectric response of a material. For dark matter which couples to electrons, this approach automatically accounts for in-medium screening effects, which were not included in previous rate calculations for semiconductor targets. We show that the screening effects appear for both scalar and vector mediators. The result is a non-negligible reduction of reach for direct detection experiments which use dielectric materials as targets. We also explore different determinations of the dielectric response, including first-principles density functional theory (DFT) calculations and a data-driven analytic approximation using a Mermin oscillator model.
\end{abstract}

\maketitle

\section{Introduction}

An increasingly diverse set of  underground direct detection experiments remains the most promising direct probe of the nature of the dark matter (DM). In a set of pioneering papers, Essig et.~al.~\cite{Essig:2011nj,Essig:2012yx,Essig:2015cda} generalized the search for DM beyond the traditional nuclear recoil paradigm, by showing that it is feasible to search for electron recoils from scattering of sub-GeV dark matter in noble liquids and semiconductors. There have since been numerous studies of DM scattering in these materials~\cite{Graham:2012su,Lee:2015qva,Ibe:2017yqa,Essig:2017kqs,Essig:2018tss,Emken:2019tni,Baxter:2019pnz,Essig:2019xkx,Catena:2019gfa,Buch:2020xyt,Andersson:2020uwc,Radick:2020qip}, as well as proposals for other targets that are sensitive to electron recoils~\cite{Hochberg:2015pha,Hochberg:2015fth,Hochberg:2016ntt,Hochberg:2017wce,Derenzo:2016fse,Kurinsky:2019pgb,Griffin:2019mvc,Blanco:2019lrf,Trickle:2019nya,Geilhufe:2019ndy,Hochberg:2019cyy,Coskuner:2019odd,Griffin:2020lgd}. Nowadays electron recoils are leveraged by every major experimental collaboration, and they are or will be a primary detection channel for experiments such SENSEI~\cite{Barak:2020fql}, DAMIC~\cite{Castello-Mor:2020jhd}, SuperCDMS~\cite{Amaral:2020ryn} and LBECA~\cite{Bernstein:2020cpc}.  For semiconductor targets in particular, the full calculation of the DM-electron scattering rate was first performed in \cite{Essig:2015cda} through an explicit calculation of the electronic wave functions with density functional theory (DFT) methods. Their calculation was recently extended to a broader range of semiconductors in~\cite{Griffin:2019mvc,Trickle:2019nya}, using a similar procedure. 

In this paper, we formulate a new approach to calculate the DM-electron scattering rate in the broad class of dielectric materials, by expressing the rate in terms of the dielectric response $\epsilon(\omega,\bfk)$. As the dielectric response is dominated by the electron response for energies $(\omega)$ above the band gap, this gives an alternate way to understand DM-electron scattering and leads to quantitatively different scattering rates, as screening effects are automatically included. Furthermore, the dielectric function is extensively studied in both condensed matter theory and experiment. Rewriting DM scattering in this way thus provides a more direct translation between quantities of interest for condensed matter and dark matter physicists. 

The dielectric response of a material determines the energy loss function (ELF), which is defined as the imaginary part of the inverse dielectric function
\begin{align}
    \Im \left[ \frac{-1}{\epsilon(\omega,\bfk)} \right].
\end{align}
This quantity describes the rate to lose momentum $\bfk$ and energy $\omega$ for a charged particle passing through the material.\footnote{Note that we will use $\epsilon(\omega,\bfk)$ throughout to mean the longitudinal dielectric function, and we work in the approximation that the dielectric function is a diagonal matrix in reciprocal lattice space. For more details, see Appendix~\ref{app:dielectric}. } The ELF is closely related to the dynamic structure factor $S(\omega,\bfk)$, which describes the rate to create density fluctuations in the medium, independently on the nature of the external probe.  We can therefore directly relate the ELF to the dark matter scattering rate. By writing the scattering rate in terms of the ELF or $S(\omega,\bfk)$, we are moreover accounting for in-medium effects such as screening, as well possible collective excitations such as plasmons \cite{Kurinsky:2020dpb,Kozaczuk:2020uzb}. The same ELF also plays an important role in the Migdal effect, or inelastic DM-nucleus scattering, which we studied in a companion paper~\cite{Knapen:2020aky}. (See also \cite{Liang:2020ryg}.)

For any scalar or vector mediator coupling to electrons, we can treat DM as an external source which couples to electron number density $n(\bfr,t)$. In linear response theory, perturbations of the electron number density in the medium can be determined by the susceptibility 
\begin{equation}\label{eq:susceptibility}
\chi(\omega,\bfk) = -\frac{i}{V} \int_0^\infty dt\, e^{i \omega t} \langle [ n_\bfk(t), n_{-\bfk}(0) ] \rangle,
\end{equation} 
given here in Fourier space, with $V$ the volume and $\bfk$ and $\omega$ respectively the momentum and energy of the perturbation. The expectation value in \eqref{eq:susceptibility} includes the thermal average. The fluctuation-dissipation theorem relates the susceptibility to the dynamical structure factor, which parametrizes the rate at which excitations are emitted or absorbed by the system: 
\begin{align}
   \Im \chi(\omega,\bfk) = -\frac{1}{2} (1 - e^{-\beta \omega}) S(\omega,\bfk).
   \label{eq:dissipation_theorem}
\end{align}
Here the dynamic structure factor is defined as
\begin{align}\label{eq:skodef}
    S(\omega,\bfk) \equiv
    \frac{2 \pi}{V} \sum_{i,f} \frac{e^{-\beta E_i}}{Z}| \langle f &| n_{-\bfk} | i \rangle |^2 \delta(\omega + E_i - E_f) 
\end{align}
with $\beta=1/k_B T$ and $Z$ the partition function of the system. Eq.~\eqref{eq:skodef} should remind the reader of Fermi's golden rule, and $S(\omega,\bfk) $ is directly proportional to the differential DM-electron scattering rate. Using the relationship between the susceptibility and dielectric response
\begin{align}
    \frac{1}{\epsilon(\omega,\bfk)} = 1 + \frac{4 \pi \alpha_{em}}{k^2} \chi(\omega,\bfk),
\end{align}
we can write the structure factor as
\begin{align}
S(\omega,\bfk)=    \frac{k^2}{2\pi \alpha_{em}} \frac{1}{1 - e^{-\beta \omega}}  \Im \left[ \frac{-1}{\epsilon_L(\omega,\bfk)} \right]. 
   \label{eq:ELF_structurefactor}
\end{align}
This relation is well known in the condensed matter literature, see e.g.~\cite{girvin_yang_2019}.

In the remainder of this paper we explore the consequences of this relationship for dark matter electron scattering. The main  difference with previous works in the literature is essentially that, writing the ELF as ${\rm Im}(\epsilon(\omega, \bfk))/| \epsilon(\omega, \bfk)|^2$, we see that a screening factor of $1/| \epsilon(\omega, \bfk)|^2$ is included inside the dynamic structure factor. Previous works studying DM scattering in semiconductors~\cite{Essig:2015cda,Griffin:2019mvc,Trickle:2019nya} primarily considered the approximation $| \epsilon(\omega, \bfk)|^2 \approx 1$. Since the DM scattering rate is dominated by $k\gtrsim$ keV, this assumption is not unreasonable, but with detailed calculations we find that screening can affect the rate by a factor of a few in Si and Ge. In addition, while the importance of accounting for screening has been well understood for vector mediators, screening for scalar-mediated scattering was only pointed out more recently in Ref.~\cite{Gelmini:2020xir} (see also Ref.~\cite{Hardy:2016kme} for discussion of in-medium effects for scalars). In this work, we put scalar and vector mediated scattering on the same footing and show how they lead to identical response functions. We also show how scattering form factors discussed in the literature relate to the dielectric response, and perform detailed calculations of the screening effect in semiconductor targets relevant for current low-threshold experiments.

In the following section, we show how the DM-electron scattering rate relates to the dynamic structure factor or ELF. In section~\ref{sec:dielectric}, we discuss different ways to determine the dielectric function and thus the ELF, including the details of our DFT calculations for semiconductors. In section~\ref{sec:DMresults} we present the implications for DM scattering in semiconductors and superconductors. We conclude in section~\ref{sec:conclusions}.

\section{DM-electron scattering as dielectric response \label{sec:DMscattering} }
The most common models which predict dark matter-electron scattering involve a scalar or vector mediator which couple respectively to the electron number density and the electron current. In the nonrelativistic limit, the leading interactions of the mediator are the same for both cases:
\begin{align}\label{eq:UVinteraction}
    &-{\cal L} \supset g_\chi \phi \bar \chi \chi + g_e \phi \bar e e &\rightarrow& \ g_\chi \phi n_\chi + g_e \phi n \nonumber\\
    -{\cal L} &\supset g_\chi V_\mu \bar \chi \gamma^\mu \chi + g_e V_\mu \bar e \gamma^\mu e  &\rightarrow& \ g_\chi V_0 n_\chi + g_e V_0 n 
\end{align}
since scattering via the $0^{\mathrm{th}}$ component of the vector dominates. Here $n_\chi$ and  $n$  are respectively DM and electron number densities.  This makes is it manifest that in the non-relativistic limit the scalar and vector mediators ought to give identical rates, up to the rescaling of the coupling constants. Note that the vector here could represent a kinematically-mixed dark photon in the interaction basis, or another vector.

Given the similarity in these interactions, we can thus consider a general mediator with coupling to electrons $g_e$ and coupling to the DM $g_\chi$. We will write the mass of the mediator as $m_V$, although it could also be a scalar. The coupling between the electron density perturbation $n_\bfk$ and the external potential to the DM is then given by 
\begin{align}\label{eq:effham}
    H_{\rm ext} =  \int \frac{d^3 \bfk}{(2 \pi)^3} n_{\bfk} \times \left( \frac{g_\chi g_e e^{i \bfk \cdot \bfx}}{k^2 + m_V^2} \right).
\end{align}
where the term in the parentheses represents the \emph{external and thus unscreened} potential due to the DM (where $\bfx$ is DM position). In this basis, all in-medium corrections will be included in $S(\omega,\bfk)$, as the propagator itself receives no corrections.  In the particle physics literature the interaction term in \eqref{eq:effham} is often written in terms of the \emph{total} potential felt by the electrons, especially so in the context of a kinetically mixed dark photon mediator. In this basis the propagator receives a multiplicative correction of the form $1/\epsilon(\omega,k)$, and one defines a different structure factor, without the screening factor. The approaches are equivalent. However by working with the external rather than the total potential, the parallel between the scalar and the vector mediator in \eqref{eq:UVinteraction} is more manifest.

Evaluating the Hamiltonian in \eqref{eq:effham} between initial and final DM states of momentum $\bfp_i$ and $\bfp_f$, respectively, as well as initial and final electron fluid states $|i \rangle, |f \rangle$, we find the matrix element
\begin{align}
    {\cal M} = \frac{g_\chi g_e}{V(k^2 + m_V^2)} \langle f |n_{-\bfk} |i \rangle \delta_{\bfp_i - \bfp_f,\bfk}
\end{align}
where in the continuum limit we can write the Kronecker delta function as a Dirac delta function, $\delta_{\bfp_i - \bfp_f,\bfk} = (2\pi)^3/V \times \delta(\bfp_i - \bfp_f - \bfk) $.
We now use Fermi's Golden rule, and sum over initial states $|i \rangle$ weighted by $e^{-\beta E_i}/Z$, as well as over final states. Inserting a factor of unity as $\int d\omega \delta(\omega + E_i - E_f)$, we obtain a DM scattering rate
\begin{align}
	R = \frac{1}{\rho_T} \frac{\rho_\chi}{m_\chi} \frac{ \pi \bar \sigma_e}{\mu_{\chi e}^2} \int d^3 v\, f_\chi(v)   \, \frac{ d^3 \bfk}{(2\pi)^3} \, d\omega \times \, \nonumber \\
	\delta\left( \omega + \frac{k^2}{2 m_\chi} - \bfk \cdot \bfv \right)  |F_{DM}(k)|^2 \, S(\omega,\bfk)
\end{align}
where $\rho_T$ is target density, $\mu_{\chi e}$ is DM-electron reduced mass, and $f_\chi(v)$ is the DM velocity distribution. Here we  used the conventional definition of DM-electron scattering cross section $\bar \sigma_e$ in terms of couplings~\cite{Essig:2015cda}:
\begin{align}
	\bar \sigma_e = \frac{\mu_{\chi e}^2 g_e^2 g_\chi^2}{\pi \big((\alpha m_e)^2 + m_V^2\big)^2}.
\end{align}
and the DM-mediator form factor is defined as
\begin{align}
    F_{DM}(k) = \frac{ m_{V}^2 + \alpha^2 m_e^2}{m_{V}^2 + k^2}.
\end{align}
Plugging in \eqref{eq:ELF_structurefactor}, we arrive at our master formula for the scattering rate
\begin{widetext}
\begin{align}\label{eq:mastereq}
	R = \frac{1}{\rho_T} \frac{\rho_\chi}{m_\chi} \frac{  \bar \sigma_e}{\mu_{\chi e}^2} \frac{\pi}{\alpha_{em}}\int\! d^3 v\, f_\chi(v)   \int\!\! \frac{ d^3 \bfk}{(2\pi)^3} k^2   |F_{DM}(k)|^2 \int\!\frac{d\omega}{2\pi}\,  \,   \frac{1}{1 - e^{-\beta \omega}} \Im \left[ \frac{-1}{\epsilon_L(\omega,\bfk)} \right]\delta\left( \omega + \frac{k^2}{2 m_\chi} - \bfk \cdot \bfv \right).
\end{align}
\end{widetext}

To compare this form of the rate with previous works in the literature, we use the Lindhard form for $\epsilon(\omega,\bfk)$. The Lindhard dielectric function, also known as the random phase approximation (RPA), is the leading-order polarization due to electron-hole excitations. It is given by~\cite{DresselGruner,Adler}
\begin{align}
    	\epsilon^{\rm RPA}(\omega,\bfk) = 1 - &\frac{4 \pi \alpha_{em}}{V k^{2}} \sum_{\bfp,\bfp',\ell, \ell'}  | \langle \bfp', \ell' |e^{i \bfk \cdot \bfr} | \bfp, \ell \rangle |^{2} \nonumber \\
    	&\times  \lim_{\eta \to 0} \frac{ f^{0}(\omega_{\bfp', \ell'}) - f^{0}(\omega_{\bfp,\ell}) }{ \omega_{\bfp', \ell'} - \omega_{\bfp, \ell} - \omega - i \eta } ,
	\label{eq:lindhard}
\end{align}
where we sum over states labeled by momentum $\bfp$ and band $\ell$. There is also an implicit sum over spin states.  The thermal occupation of the electron state with energy $\omega_{\bfp, \ell}$ is $f^0(\omega_{\bfp, \ell}) =1/\left[\exp(\beta(\omega_{\bfp, \ell} - E_F)) +1 \right]$, where $E_F$ is the Fermi energy.
Using \eqref{eq:lindhard} in \eqref{eq:ELF_structurefactor}, we find
\begin{align}
\label{eq:structurefactor_RPA}
    S(\omega,& \bfk) =
    \frac{2 \pi}{V |\epsilon^{\rm RPA}(\omega,\bfk)|^2 } \sum_{\bfp, \bfp' \newline \ell, \ell'} | \langle \bfp, \ell' | e^{i \bfk \cdot \bfr} | \bfp, \ell \rangle |^2  \\
     & \times  f^{0}(\omega_{\bfp, \ell}) ( 1 - f^{0}(\omega_{\bfp', \ell'}) ) \delta(\omega + \omega_{\bfp,\ell} - \omega_{\bfp',\ell'}). \nonumber
\end{align}
Here we recognize a rate to create single-electron excitations, but with a screening factor of $1/| \epsilon(\omega, \bfk)|^2$. Many works have considered this screening effect for vector mediators, by defining an effective coupling in the medium. Here we show that it should apply to scalars too, and include it inside the dynamic structure factor. 

Our formulation of DM-electron scattering in terms of a structure factor is then identical to that of Refs.~\cite{Griffin:2019mvc,Trickle:2019nya} when the dielectric function is computed in RPA, with the only difference being the screening factor appearing in $S(\omega,\bfk)$. Similarly, our results are equivalent to those of Ref.~\cite{Essig:2015cda} when the RPA dielectric function is used and $| \epsilon(\omega, \bfk)|^2 \to 1$. More explicitly, we find that the crystal form factor of Ref.~\cite{Essig:2015cda} is given by
\begin{align}
    |f_{\rm crystal}(k,\omega)|^{2} = \frac{k^{5} V_{\rm cell}}{8 \pi^{2} \alpha_{em}^{2}m_{e}^{2}}  \Im ( \epsilon^{\rm RPA}(\omega, \bfk)) . \label{eq:fcrystal_vs_Imeps}
\end{align}
with $V_{\rm cell}$ the volume of the unit cell. The relationships between the different conventions for DM-electron scattering are discussed in more detail in Appendix~\ref{app:formalismcomparison}.

\section{Dielectric response \label{sec:dielectric}}

In this section, we discuss two approaches for determining $\epsilon(\omega,\bfk)$: a data-driven analytic approximation, and density functional theory (DFT) calculations. The various calculations of $\epsilon(\omega,\bfk)$ are compared with each other and with experimental data. Readers who are interested primarily in the DM scattering reach can proceed directly to Sec.~\ref{sec:DMresults}.

\subsection{Mermin oscillator model\label{sec:mermin}}

A semi-analytic approximation to $\epsilon(\omega,\bfk)$ is valuable to quickly obtain results for many materials, as compared to numerically expensive DFT calculations. Direct measurements of $\epsilon(\omega,\bfk)$ at several $\omega, \bfk$ points are moreover often available to anchor such a  semi-analytic description. To fully make use of these measurements, a self-consistent interpolation is however needed which preserves the various sum rules and symmetries associated with $\epsilon(\omega,\bfk)$. For some materials the available data is also restricted to the optical ($\bfk=0$) limit, and a well-motivated extrapolation to finite $\bfk$ is therefore desirable. 

One of the simplest, analytic models of $\epsilon(\omega,\bfk)$ is the Lindhard model for a homogeneous electron gas, for which the dielectric function can be characterized entirely by its Fermi velocity $v_F = k_F/m_e$ and plasma frequency $\omega_p = \sqrt{4 \pi \alpha_{em} n_e/m_e}$, with $n_e$ the electron number density. The dielectric function is also isotropic in $\bfk$ in this case, and can be directly evaluated in \eqref{eq:lindhard} by inserting plane wave states. The result is \cite{DresselGruner}
\begin{align}\label{eq:eps_lindhard_gas}
    \epsilon_{\rm Lin}(\omega,k;\omega_p)=&1+\frac{3\omega_p^2}{k^2 v_F^2}\Bigg[f_1\left(\frac{\omega}{k v_F},\frac{k}{2 m v_F}\right)\nonumber\\
    &+i f_2\left(\frac{\omega}{k v_F},\frac{k}{2 m v_F}\right)\Bigg]
 \end{align}
with
\begin{align}
v_F&=\left(\frac{3\pi \omega_p^2}{4 \alpha_{em} m_e^2}\right)^{1/3}\nonumber\\
f_1(u,z)&=\frac{1}{2}+\frac{1}{8z}\left[g(z-u)+g(z+u)\right]\nonumber\\
f_2(u,z)&=\left\{\begin{array}{l}
\frac{\pi}{2}u,\quad z+u<1\\
\frac{\pi}{8z}\left(1-(z-u)^2\right),\quad |z-u|<1<z+u\\
0,\quad |z-u|>1
\end{array}\right.\nonumber\\
g(x)&=(1-x^2)\log\left|\frac{1+x}{1-x}\right|
\end{align}
where we have explicitly separated the results into its real and imaginary parts. The main shortcoming of the Lindhard model is that the plasmon peak has zero width, which is certainly not the case in semiconductors such as Si and Ge. This problem is addressed in the Mermin model \cite{PhysRevB.1.2362} 
\begin{equation}
    \epsilon_{\text{Mer}}(k,\omega;\omega_p,\Gamma) = 1+\frac{(1+i \Gamma/\omega)(\epsilon_{\text{Lin}}(k,\omega+i\Gamma)-1)}{1+(i\Gamma/\omega)\frac{\epsilon_{\text{Lin}}(k,\omega+i\Gamma)-1}{\epsilon_{\text{Lin}}(k,0)-1}}
\end{equation}
with $\Gamma$ the width of the plasmon pole. By construction $\epsilon_{\text{Mer}}(k,\omega;\omega_p,0)=\epsilon_{\text{Lin}}(k,\omega;\omega_p)$. $\epsilon_{\text{Mer}}$ is moreover designed such that the various sum rules on the the dielectric function are explicitly satisfied. 

Both the Mermin and Lindhard models however apply to a homogeneous electron gas, which is a far cry from a realistic material. This is often addressed in a phenomenological way by modeling the material as a superposition of many electron gas clouds with different densities. In other words, one describes ELF as a linear combination of Mermin dielectric functions. Here we follow the procedure outlined in \cite{PhysRevA.58.357,VOS2019242}
\begin{align}
\text{Im}\left[\frac{-1}{\epsilon(\omega,k)}\right]=&\sum_i A_i(k) \text{Im}\left[\frac{-1}{\epsilon_{\text{Mer}}(\omega,k;\omega_{p,i},\Gamma_i)}\right]\nonumber\\
&\times\theta(\omega-\omega_{edge,i})
\end{align}
with
\begin{equation}
A_i(k)=A_i(0)\frac{\int_0^\infty\!\! d\omega\, \omega\, \text{Im}\left[\frac{1}{\epsilon_{\text{Mer}}(\omega,0;\omega_{p,i},\Gamma_i)}\right]\theta(\omega-\omega_{edge,i})}{\int_0^\infty\!\! d\omega\, \omega \, \text{Im}\left[\frac{1}{\epsilon_{\text{Mer}}(\omega,k;\omega_{p,i},\Gamma_i)}\right]\theta(\omega-\omega_{edge,i})}
\end{equation}
where the $\omega_{p,i}$, $\Gamma_i$, $\omega_{edge,i}$ and $A_i(0)$ are fitted to experimental data. One can use as many Mermin oscillators as needed to describe the experimental data. The real part of $1/\epsilon(\omega,k)$ can be obtained through a Kramers-Kr\"onig transformation. This approach also makes it possible to include the semi-core electrons in a phenomenological manner \cite{VOS2019242}, something which is computationally difficult to do in first principles DFT calculations.

\begin{figure*}[t]
\includegraphics[width=\textwidth]{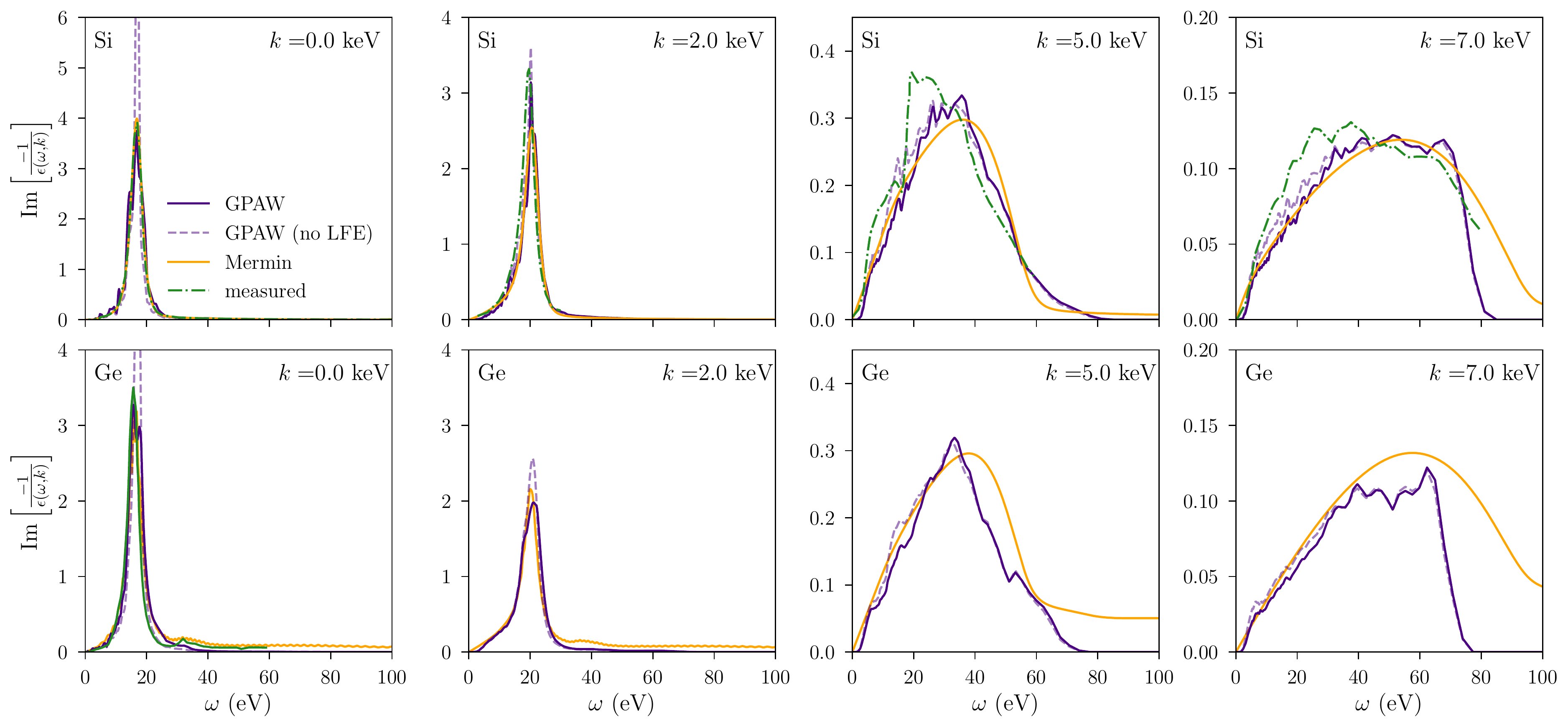}
\caption{The ELF evaluated with \texttt{GPAW} and the Mermin oscillator method, as implemented in \texttt{chapidif}. When a measurement is available, it is overlaid as well. For $k=0$ the Si and Ge data are taken from respectively \cite{Optical} and \cite{NOVAK20087}. At finite $k$ for Si the measured ELF is taken from the Weissker et~al.~dataset \cite{Weissker}. The Mermin oscillators were fit to optical ($k=0$) data \cite{NOVAK20087,Dingdatabase}; the Weissker et~al.~data for finite $k$ values is independent and not included in this fit. The discrepancy at high $\omega$ is due to the fact the \texttt{GPAW} calculation only includes the lowest 70 bands in computing the ELF, and hence does not capture the dielectric response for $\omega \gtrsim 70$ eV. Note that for DM with maximum speed of $\sim 750$ km/s,  only the phase space with $k \gtrsim 4\, \keV\times \omega/(10\, {\rm eV})$ contributes to DM-electron scattering. The sharp plasmon resonance in the first two columns therefore does not contribute to the scattering rate. Those panels are meant only as a validation of our methods. } \label{fig:epscomparison}
\end{figure*}

For our calculations we make use of the \texttt{chapidif} package\footnote{We thank Maarten Vos for providing us with a $\beta$-version of the code.} \cite{chapidif}, with experimental inputs taken from \cite{NOVAK20087,Dingdatabase}, all obtained in the optical ($k=0$) limit. In Fig.~\ref{fig:epscomparison} we show ELF for Si and Ge, and compare with a DFT calculation with the \texttt{GPAW} code (see next section). For Si, we also compare with the finite-$k$ data from Weissker et.~al.~\cite{Weissker}, which is independent from the data used to fit to the Mermin model. We find good agreement between all three methods, except for the high-$k$, high-$\omega$ regime. Both the DFT and Mermin oscillator methods suffer from increased uncertainties in this regime: in the DFT calculation, higher values of $\omega$ require more bands to be included, which increases the computational complexity. With the Mermin oscillator method the uncertainties are expected to grow the further one deviates from the optical limit. This may be addressed by including finite-$k$ data in the fit. For our numerical calculations of the DM scattering rate in semiconductors we will rely on the DFT method, and reserve a more detailed comparison to Appendix~\ref{sec:comparison}.

\subsection{DFT calculations\label{sec:DFT}}

As an alternative to the phenomenological approach of the previous subsection, it is also possible to determine the dielectric response of a material from first principles. In contrast to the case of a homogeneous electron gas, in a crystal,  response functions are only invariant under lattice periodicity, so that in momentum space the full dielectric function is written as a matrix $\epsilon_{\bfG \bfG'}(\bfq)$ where $\bfq$ is restricted to the first Brillouin zone (1BZ) and $\bfG, \bfG'$ are reciprocal lattice vectors. The dielectric function is then treated as a matrix in reciprocal lattice vectors. Here, we provide an overview of microphysical calculations of this dielectric response in the framework of time-dependent density functional theory (TDDFT), while additional details are reviewed in Appendix~\ref{app:dielectric}.

In the TDDFT approach, one maps the system of interacting electrons in the presence of an external (time-dependent) potential to a system of non-interacting electrons in the presence of an effective potential. The latter is known as the Kohn-Sham (KS) system~\cite{PhysRev.140.A1133} and is much simpler to work with, since one has only to deal with effective single electron wavefunctions. Quantities such as the susceptibility $\chi_{\mathbf{G}\mathbf{G}'}(\mathbf{q},\omega)$ and the polarizability $P_{\mathbf{G}\mathbf{G}'}(\mathbf{q},\omega)$ can be related to their counterparts computed in the simpler Kohn-Sham system by requiring the change in charge density in response to a small change in the external potential (in the full system) and the effective potential (in the KS system) to be the same.

We are ultimately interested in the microscopic dielectric function, which is related to the polarizability by~\cite{thesis}
\begin{equation} \label{eq:eps_DFT}
\epsilon_{\bfG\bfG'}(\bfq,\omega) = \delta_{\bfG \bfG'} - \frac{4\pi \alpha_{em}}{\left| \bfq+\bfG \right| \left|\bfq+\bfG' \right|} P_{\bfG\bfG'}(\bfq,\omega).
\end{equation}
In the random phase approximation, the polarizability is approximated with the KS susceptibility $P_{\bfG\bfG'}(\bfq,\omega) \approx \chi^{KS}_{\bfG,\bfG'}(\bfq,\omega)$ (see Appendix~\ref{app:dielectric}), and thus
\begin{equation}
\epsilon_{\bfG\bfG'}(\bfq,\omega) \simeq \delta_{\bfG \bfG'} - \frac{4\pi \alpha_{em}}{\left| \bfq+\bfG \right| \left|\bfq+\bfG' \right|} \chi^{KS}_{\bfG\bfG'}(\bfq,\omega) \label{eq:eps_DFT_appx}
\end{equation}
which, neglecting the off-diagonal pieces, is simply the Lindhard dielectric function of Eq.~\eqref{eq:lindhard} computed with KS wavefunctions and extended to momenta $\bfk = \bfq+\bfG$ outside the 1BZ (see also Eq.~\eqref{eq:Lindard_general}). By solving for the susceptibility in the relatively simple KS system, one arrives at an approximation for the full microscopic dielectric function. 

There exist several DFT tools to compute the KS susceptibility, and hence the RPA dielectric response. We use the public code \texttt{GPAW}~\cite{GPAW1, GPAW2} for this purpose and focus on Si and Ge semiconductors. First, the KS wavefunctions are computed. This is done at zero temperature, using a plane-wave basis with a cutoff of $E_{\rm cut} = 500$ eV, corresponding to $|\bfk|\lesssim 22$ keV. The Brillouin zone is sampled using a gamma-centered Monkhorst-Pack grid with $8 \times 8 \times 8$ $k$ points for Si, while for Ge we use a $12\times 12 \times 12$ grid. The finer grid for Ge was chosen to improve convergence of the results with respect to the grid spacing. Seventy bands are included for each spin. The KS wavefunctions are computed using the TB09 exchange-correlation functional~\cite{TB09}, and a scissor correction is applied to
match the experimentally measured Si and Ge bandgaps at $T=0$.  Note that the $3d$ electrons in Ge are treated as part of the frozen core, in contrast to e.g.~\cite{Essig:2015cda}.

\begin{figure*}[t]
\includegraphics[width=0.48\textwidth]{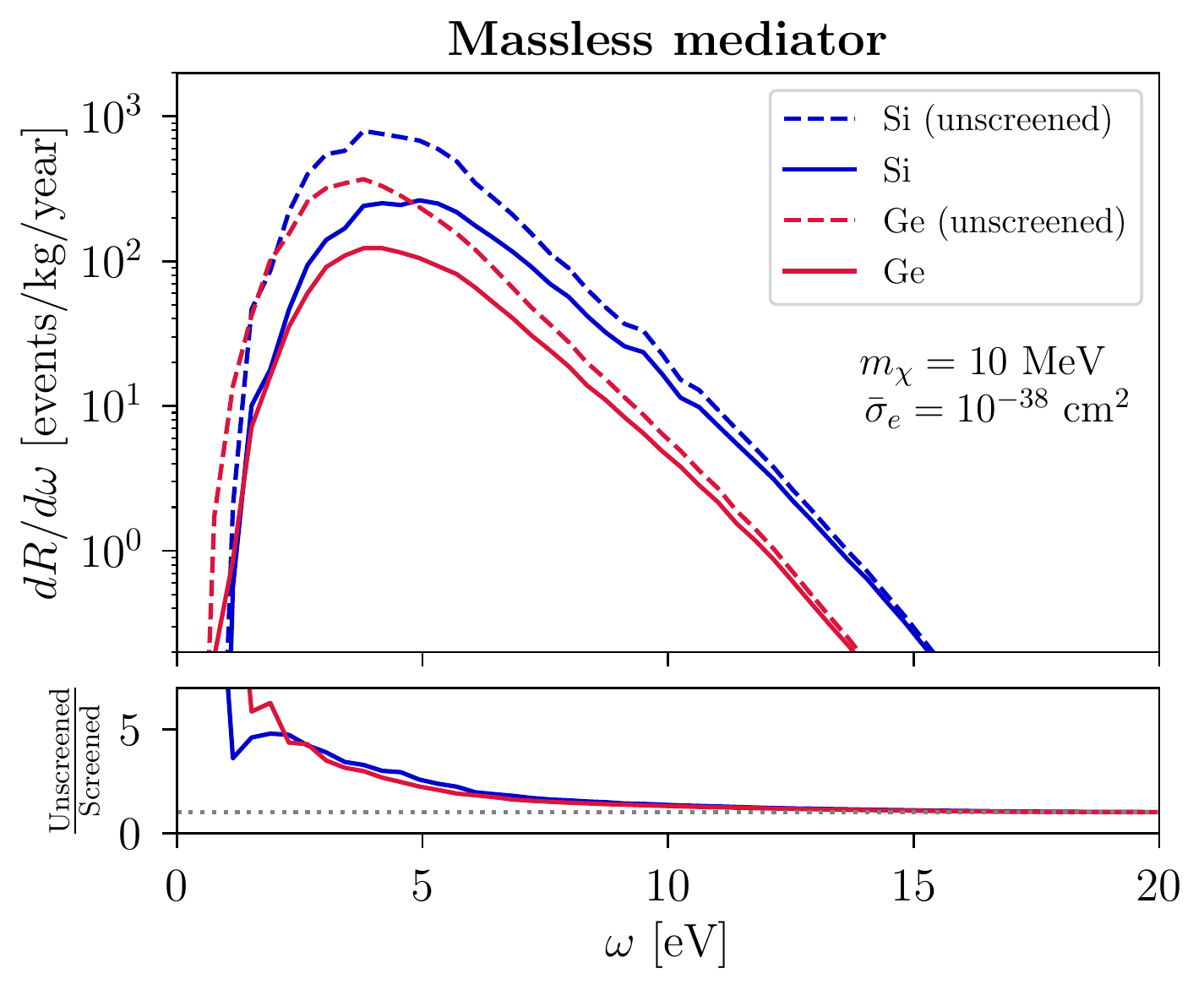}
\includegraphics[width=0.48\textwidth]{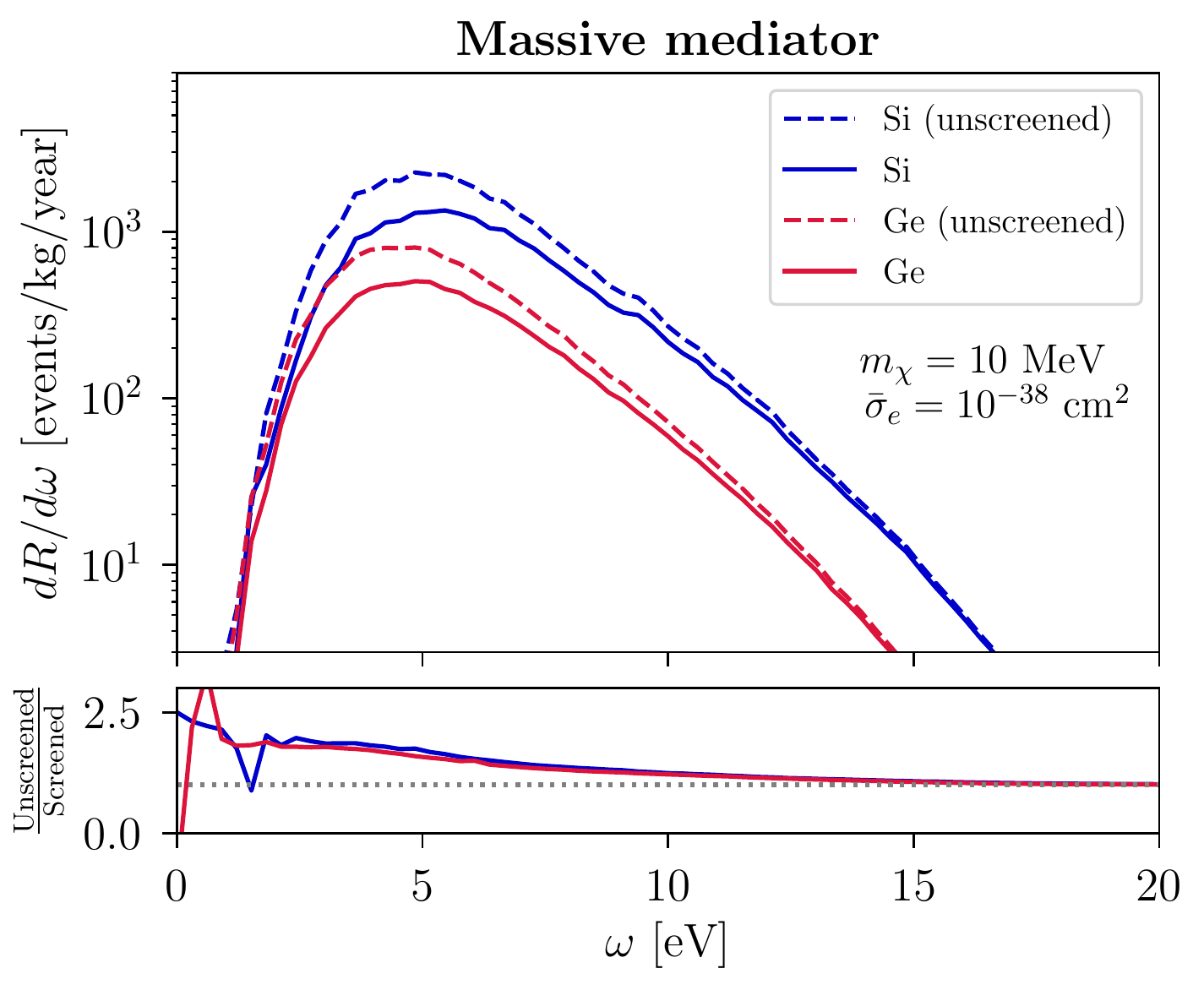}
\caption{Effect of screening on differential rate spectrum in Si and Ge semiconductors, for an example DM mass of 10 MeV and cross section $\bar \sigma_e = 10^{-38}$ cm$^2$. The bottom panel shows the ratio of the unscreened rate over the screened rate. \label{fig:screening_spectrum}}
\end{figure*}

Next, the longitudinal dielectric matrix is computed in the RPA using \eqref{eq:eps_DFT_appx} for all $\bfq \in$ 1BZ sampled by the Monkhorst-Pack grid. We will work in an approximation where we neglect the directional dependence of the response, as well as the off-diagonal components of the dielectric matrix. To this end, we define an angular-averaged dielectric function 
 \begin{equation}
\overline \epsilon (\omega, k) \equiv \frac{1}{N(k)} \sum_{\bfq, \bfG} \epsilon_{\bfG\bfG}(\omega,\bfq) \delta_{k,|\bfq+\bfG|}
\end{equation}
where $N(k)\equiv \sum_{\bfq, \bfG}  \delta_{k,|\bfq+\bfG|}$, the $\bfq$ sum runs over all 1BZ points sampled by the Monkhorst-Pack grid and and the $\bfG$ sum over all reciprocal lattice vectors up to the plane wave cutoff momentum. This quantity can then be used as an approximation to the full dielectric function in the ELF, Im$(-1/\epsilon(\omega,\bfk))\simeq \text{Im}(-1/\overline \epsilon(\omega,k))$. This approach neglects so-called ``local field effects'' (LFEs), since the off-diagonal components of the dielectric matrix are dropped altogether.

In practice, some information about the off-diagonal components of the dielectric matrix can be included by replacing $\epsilon_{\bfG\bfG} \to 1/(\epsilon^{-1}_{\bfG\bfG})$. This is known as the ``inclusion of LFEs'' in the literature. Using this quantity in the ELF results in a better fit to experiment~(see e.g.~\cite{Weissker} and Fig.~\ref{fig:epscomparison}), since at low momentum transfer this procedure amounts to averaging out the effects of the off-diagonal components of the dielectric matrix~\cite{thesis}. Approximating the loss function with $\overline \epsilon(\omega,k)$ with or without LFEs does not make a substantial difference in the experimental sensitivity to DM-electron scattering presented in the next section. We include local field effects except where stated otherwise, so that the loss function predicted by \texttt{GPAW} more closely matches experimental results.

The results computed by \texttt{GPAW} for the ELF in Si and Ge are illustrated in Fig.~\ref{fig:epscomparison} for various values of $k$. We see that generally the DFT results agree well with both experimental results (where available) and the Mermin approach described in the previous subsection. The discrepancies at large $\omega$ are due to the fact that \texttt{GPAW} only includes the lowest 70 bands in computing the loss function, so does not yield reliable results above $\sim 70$ eV for Si and Ge. 

\section{Implications for DM-electron scattering \label{sec:DMresults}}

To show the impact of screening, we now evaluate the scattering rate in example dielectric materials. We will consider the `massless mediator' limit where $m_V \ll \alpha m_e$ with $F_{DM}(k) = (\alpha m_e)^2/k^2$ and the `massive mediator' limit where $m_V \gg \alpha m_e$ with $F_{DM}(k) = 1$. As discussed before, the results here apply for both vector and scalar mediators.

Our main results focus on Si and Ge semiconductors, which are used in a number of direct detection experiments. We use $\epsilon(\omega,\bfk)$ computed in the DFT framework as described in the previous section, taking as our default the RPA dielectric function including local field effects. Again, there is only a small difference in rate whether local field effects are included or not, and we show an explicit comparison in Fig.~\ref{fig:reachcompare} in Appendix~\ref{app:formalismcomparison}. The Mermin oscillator determination of $\epsilon(\omega,\bfk)$ also gives comparable results as long as we do not consider $\omega$ too close to the band gap, which is reasonable for background-limited experiments. The results with the Mermin oscillator method are given in Appendix~\ref{sec:comparison}.
For the DM velocity distribution, we assume the Standard Halo Model with $v_{\rm esc} = 500$ km/s, velocity dispersion $v_0 = 220$ km/s, and Earth velocity $v_e = 240$ km/s. 

\begin{figure*}[t]
\includegraphics[width=0.48\textwidth]{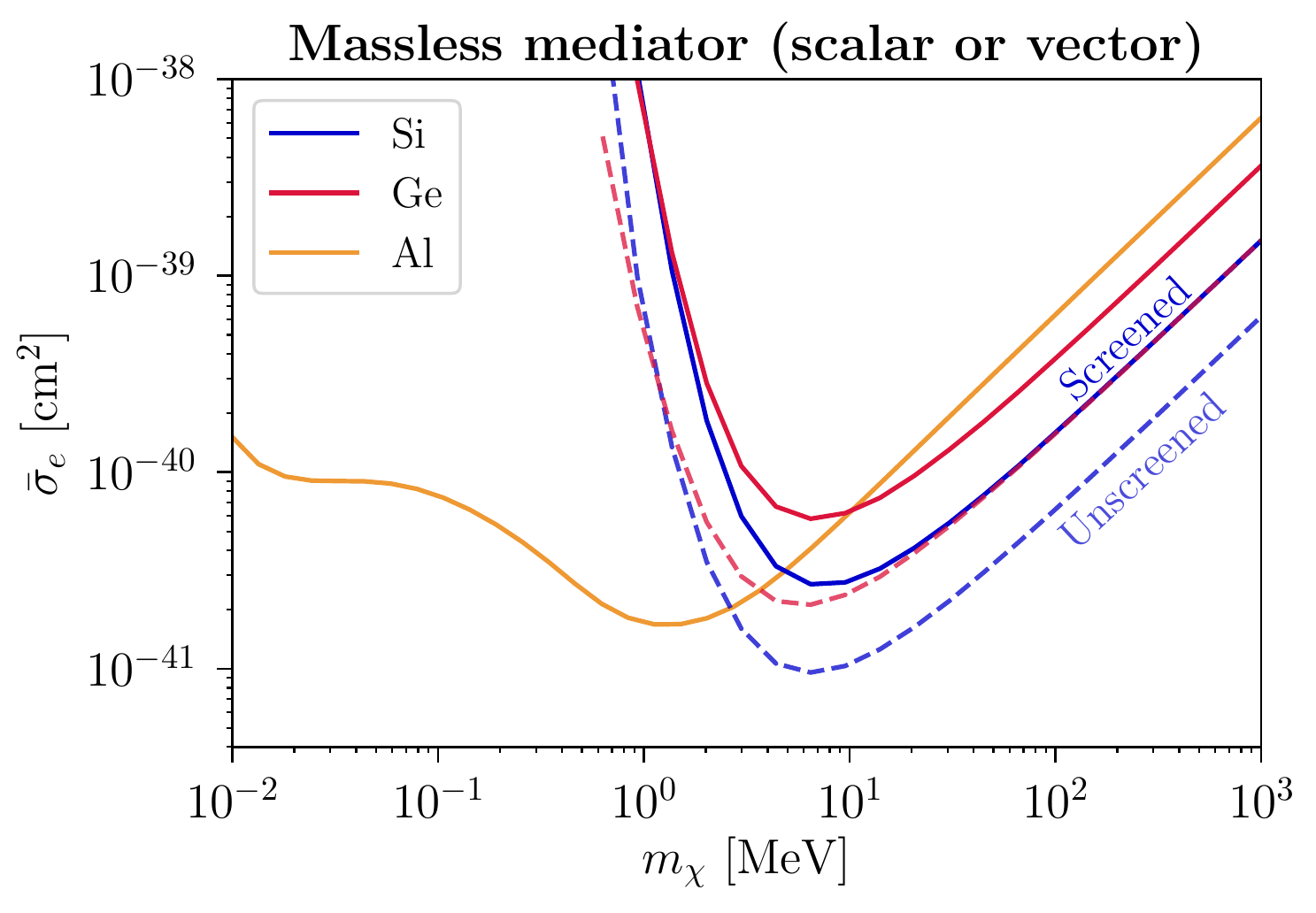}
\includegraphics[width=0.48\textwidth]{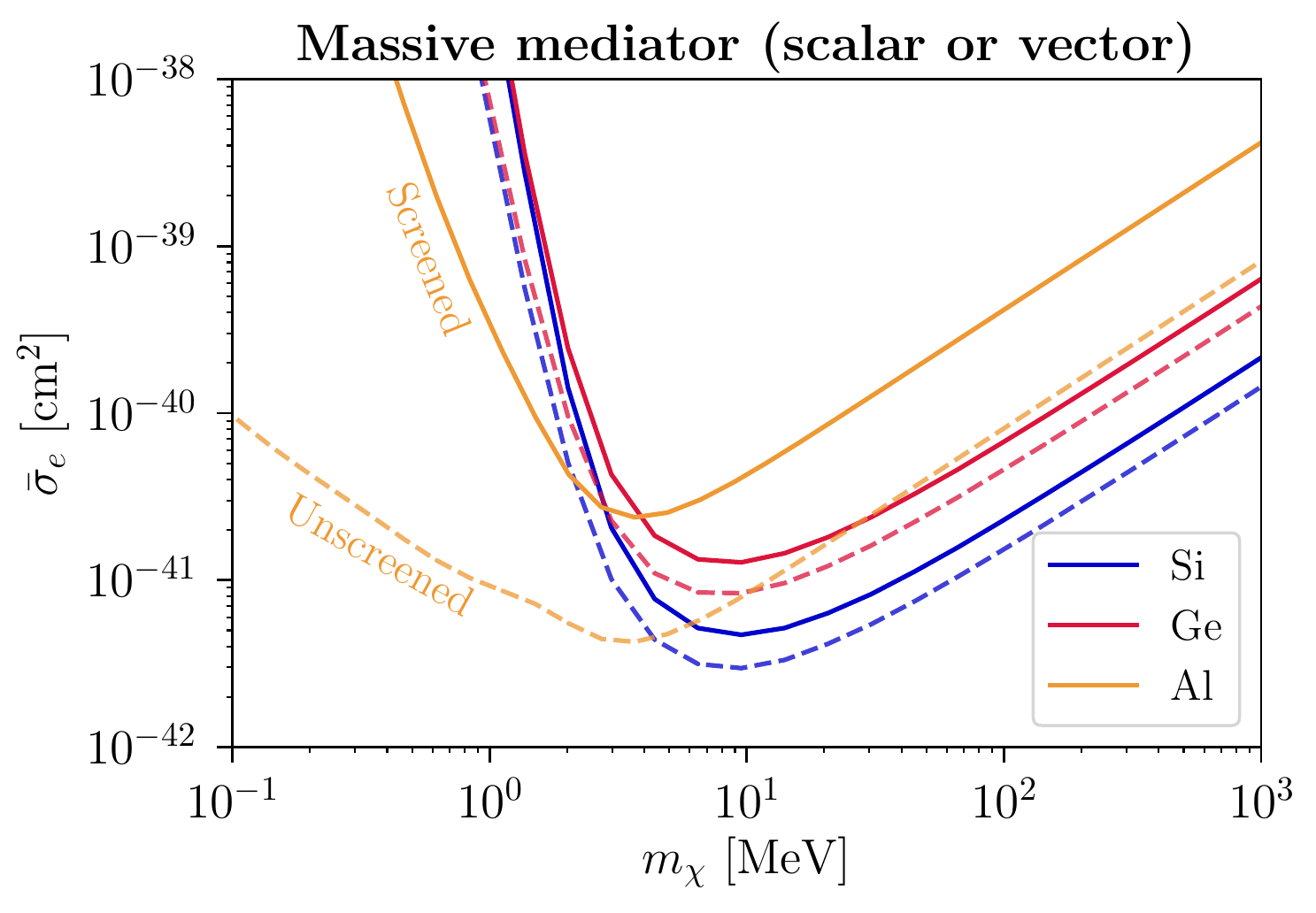}
\caption{Comparison of cross section sensitivity. The solid lines show the 95\% CL reach with kg-yr exposure for scalar or vector mediated interactions, and account for screening. The dashed lines show the reach if the screening is not included. Following the standard convention, we assume zero background down to single electron sensitivity for Si and Ge. For the Al lines, we assume an energy range of 10 meV $<\omega<$1 eV, and also zero background. In the left panel, the unscreened Al reach is many orders of magnitude stronger and is not shown on the plot. \label{fig:screening_reach}  }
\end{figure*}

\begin{figure*}[t]
\includegraphics[width=0.48\textwidth]{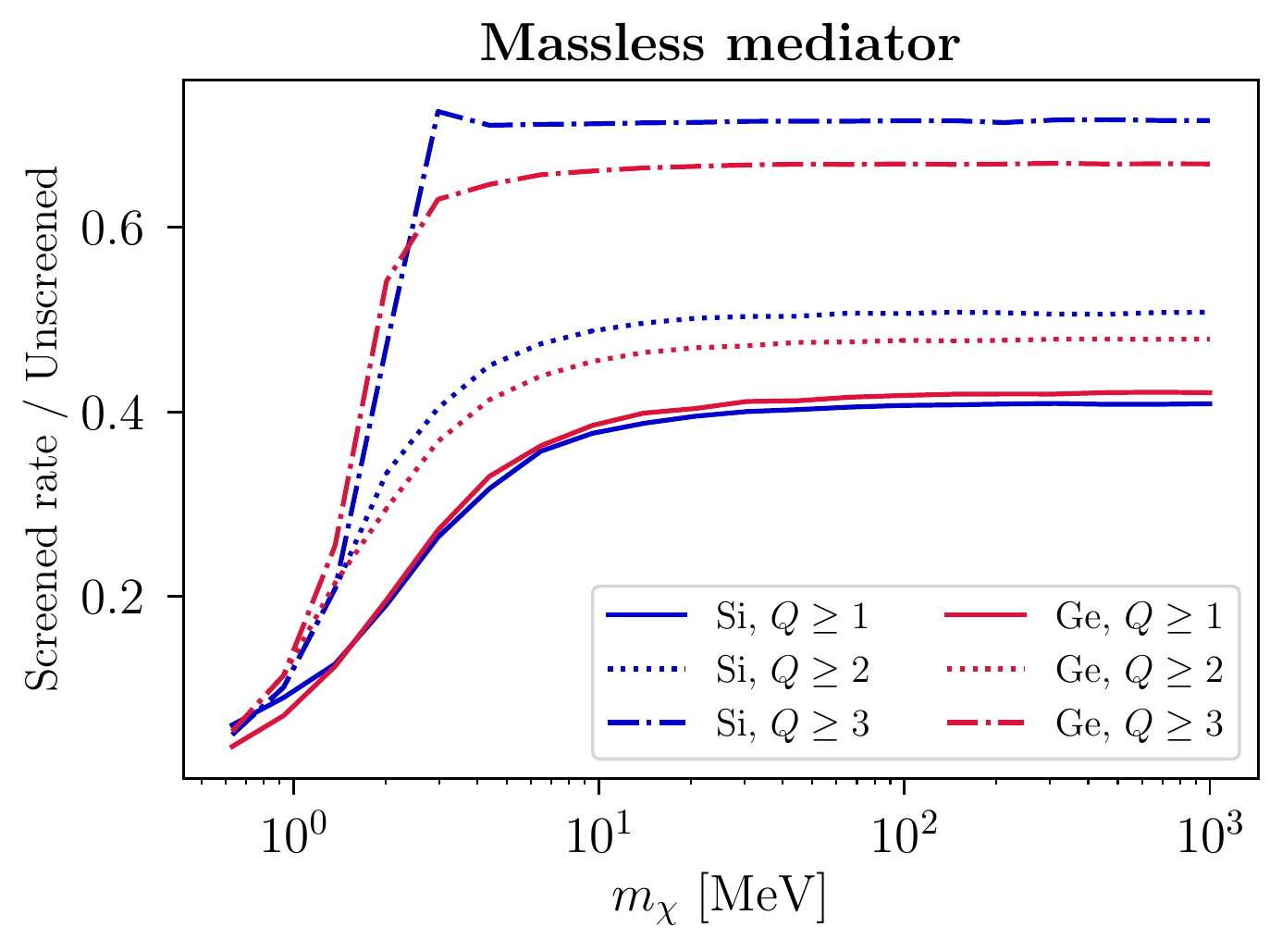}
\includegraphics[width=0.48\textwidth]{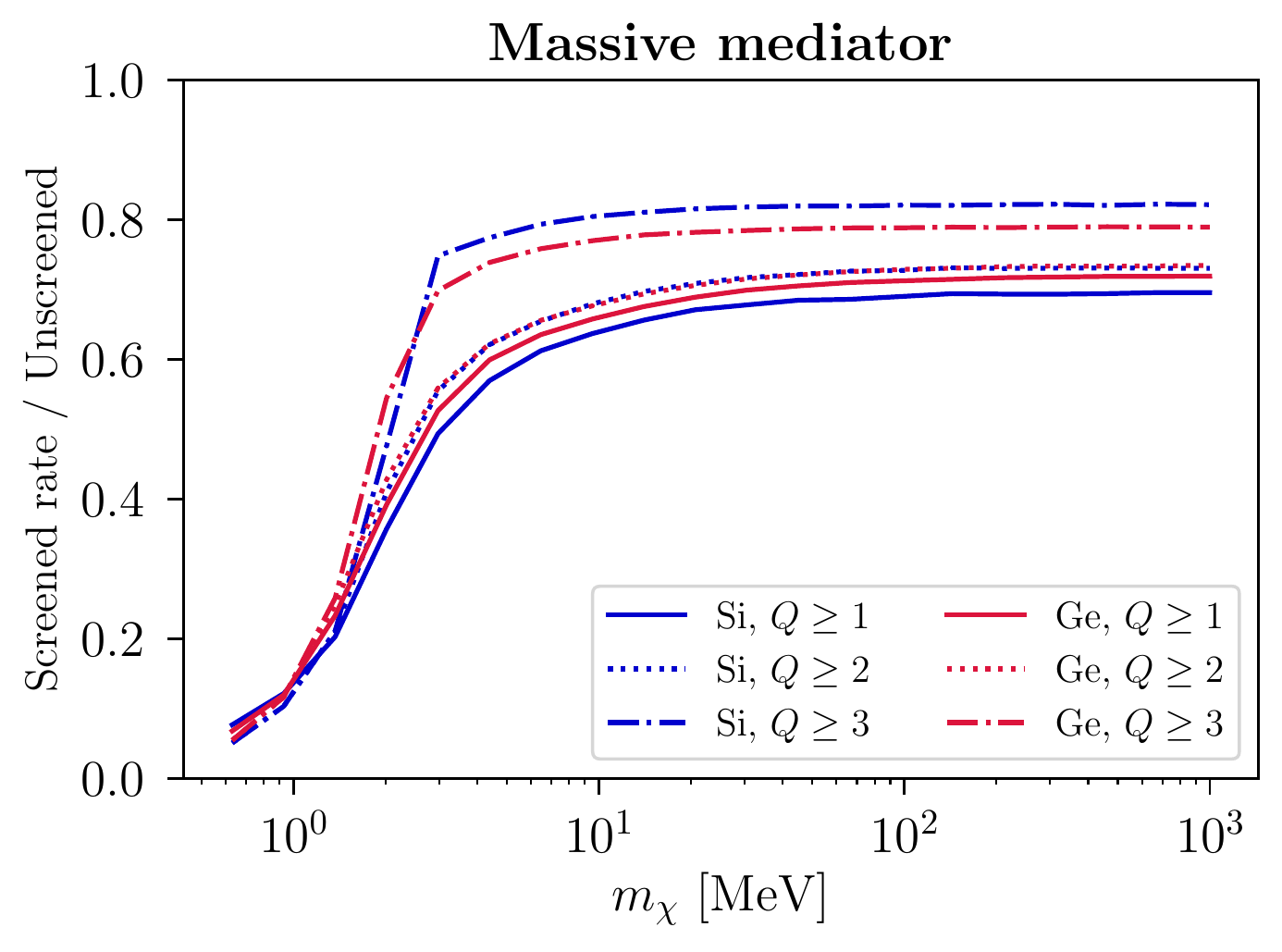}
\caption{Ratio of the screened rate to the unscreened rate, for different thresholds corresponding to 1, 2 and 3 electrons. We use $Q = 1 + \lfloor (\omega-E_g)/\varepsilon \rfloor$ where for Si $E_g = 1.11$ eV, $\varepsilon = 3.6$ eV and for Ge $E_g = 0.67$ eV, $\varepsilon = 2.9$ eV, following Ref.~\cite{Essig:2015cda}. \label{fig:threshold_dependence}}
\end{figure*}

Fig.~\ref{fig:screening_spectrum} shows the impact of screening on the differential rate spectrum, for an example DM mass of 10 MeV. Here the unscreened rate (dashed lines) is obtained by writing the ELF as $\Im(\epsilon(\omega,\bfk))/|\epsilon(\omega,\bfk)|^2$ and taking $|\epsilon(\omega,\bfk)|^2 \to 1$. The screening effects are most noticeable for lower energy deposition $\omega$, since in that case there is a larger contribution from lower momentum transfers where the screening is largest. Scattering at large $\omega$ is dominated by large $k$, with negligible screening. Similarly, we see that the effect of screening is larger for the massless mediator case, since the DM form factor $F_{DM}(k)$ enhances the rate from lower $k$ values. 

We show the corresponding effect on the DM mass and cross section reach in Fig.~\ref{fig:screening_reach}. The solid lines show the reach for scalar and vector mediators, accounting for screening effects. We assume kg-year exposure, zero background, and 95\% CL projected reach to match with the convention in the literature. The threshold is set by the electron band gap. For  $m_\chi \gtrsim 10$ MeV, there is roughly a factor of (1.4) 2.5 suppression in the total rate for (massive) massless mediators. The ratio becomes larger near threshold in $m_\chi$, since for those points the rate is restricted to $\omega$ near the band gap, where screening is more important. The screening effect is therefore reduced somewhat with higher thresholds in $\omega$, as shown in Fig.~\ref{fig:threshold_dependence}. For instance, the threshold to detect 2 electron-hole pairs is roughly 4.7 eV (3.6 eV) in Si (Ge). Setting this as the threshold, we find a screening suppression instead of 2--2.1 for massless mediators and $m_\chi \gtrsim 10$ MeV.  For massive mediators the dependence on the energy threshold is smaller.

The $\mathcal{O}(1)$ screening effects we find for Si and Ge align with our expectations for semiconductors with eV-scale electron band gaps, and it is therefore interesting to compare with a lower gap material where the screening is much stronger. We also show in Fig.~\ref{fig:screening_reach} the reach in a metal, taking Al as an example. Such targets have been proposed to be used in their superconducting phase as low-threshold dark matter detectors~\cite{Hochberg:2015fth,Hochberg:2015pha,Hochberg:2016ajh,Hochberg:2019cyy}. We thus consider sensitivity to electron recoils in the energy range 10 meV -- 1 eV, such that the material can still be approximated with the dielectric response of a metal. Here we use the Mermin oscillator method with the Al data from \cite{Dingdatabase}.  For $\omega>10$ meV, we find that the rates are in good agreement with those obtained with the Lindhard dielectric function for a free electron gas in \eqref{eq:eps_lindhard_gas}, taking $\omega_p=15$ eV. For $\omega<10$ meV the agreement between the methods is not as good, for reasons to be understood further. Out of an abundance of caution we therefore impose a \mbox{$\omega>10$ meV} threshold in Fig.~\ref{fig:screening_reach}. For massive mediators, the screening strongly limits the sensitivity to sub-MeV dark matter despite the lower thresholds. For massless mediators, in the absence of screening there is enhanced scattering with low $k$ and lower thresholds, and the unscreened reach is many orders of magnitude below what is shown on the plot. Accounting for screening, we find that there is still substantial reach to sub-MeV dark matter scattering via a massless mediator.  Thus, even with the large screening, such a low gap target could be sensitive to cosmologically interesting sub-MeV dark matter models such as that of freeze-in through a kinetically-mixed dark photon~\cite{Essig:2015cda,Dvorkin:2019zdi,Dvorkin:2020xga}.

\section{Conclusions\label{sec:conclusions}}

By considering the linear response of a dielectric material, we have shown that the differential DM-electron scattering rate in a dielectric is proportional to the energy loss function $\text{Im}[-1/\epsilon(\omega,k)]$ (see~\eqref{eq:mastereq}), which contains all relevant many-body effects associated with the target material.
The ELF is moreover very well studied theoretically and experimentally in the materials science literature, and thus provides a convenient way of mapping the detailed properties of the target material onto sensitivity estimates or limits for DM direct detection experiments. In particular, we find that screening effects need to be accounted for, both for scalar and vector mediators, which reduces the reach of any direct detection experiment with a dielectric target. We computed the ELF for Si and Ge using a first principles DFT calculation and using a data-driven, phenomenological model. Both methods broadly agree within their regime of validity. Using these results, we can quantify the importance of the screening effect in Si and Ge (see Fig.~\ref{fig:threshold_dependence}).
There are a number of possible future directions to pursue, such as accounting for angular dependence in the dielectric response for semiconductors, and applying our methodology to a broader range of materials, including others already proposed for the direct detection of electron recoils.

\acknowledgments
We thank Diego Redigolo for collaboration in early stages of this work and for useful discussions. We also thank Yonit Hochberg, Yoni Kahn, and Noah Kurinsky for useful discussions. We thank Maarten Vos for providing us with a $\beta$-version of his \texttt{chapidif} package and for his assistance with its usage and the interpretation of the results. TL is supported by the Department of Energy under grant DE-SC0019195 and a UC Hellman fellowship. JK is supported by the Department of Energy under grants DE-SC0019195 and DE-SC0009919.

\FloatBarrier
\appendix

\section{Dielectric response in a crystal \label{app:dielectric} }

In our analysis we considered the scalar longitudinal dielectric function. The more general quantity in a crystal is a dielectric tensor that is a matrix both in spatial indices and in reciprocal lattice vectors. For completeness, here we introduce the dielectric tensor and detail the approximations made in the main text.

The dielectric tensor describes the electrical response of a system to an external electric field, $\bEext$. In terms of microscopic quantities, the relationship between the external and total electric fields is
\beq \label{eq:Eext}
E_i(\om,\br)=\sum_j\int d^3 r^\prime \epsilon_{ij}^{-1}(\omega,\br,\br^\prime)E^{\rm ext}_{ j}(\omega,\br^\prime),
\eeq
which serves as the definition of $\epsilon_{ij}^{-1}(\omega,\br,\br^\prime)$. The Latin subscripts correspond to spatial indices. The fields can be written in terms of their Fourier components
\beq
\bE(\om,\br)=\frac{1}{\sqrt{V}}\sum_{\bq \in 1\text{BZ}}\, \sum_{\bG} \bE(\om,\bq+\bG)e^{i (\bq+\bG)\cdot \br}
\eeq
and similarly for $\bEext$. In the above expression, $\bq$ lies in the first Brillouin Zone ($1\text{BZ}$) and $\bG$, $\bGp$ are reciprocal lattice vectors. The dielectric tensor can be Fourier transformed as
\begin{align}\label{eq:epsFourier}
\epsilon_{ij}(\om,\br,\br^\prime)
=&\frac{1}{V}\sum_{\bq \in 1\text{BZ}}\sum_{\bG,\bGp}e^{i(\bq+\bG)\cdot \mathbf{r}}\nonumber\\
&\times \epsilon^{-1}_{ij}(\om,\bq+\bG,\bq+\bGp) e^{-i(\bq+\bGp)\cdot \mathbf{r}^{\prime}}
\end{align}
where we have used the fact the microscopic electronic response of a crystal is invariant under translations by a lattice vector $\bR$, so $\epsilon_{ij}(\om,\br,\br^\prime)=\epsilon_{ij}(\om,\br+\bR,\br^\prime+\bR)$. In Fourier space \eqref{eq:Eext} is then 
\begin{align}\label{eq:full_response}
E_i(\om,\bq+\bG)=&\sum_{\bGp}\epsilon^{-1}_{ij}(\omega,\bq+\bG,\bq+\bGp)\nonumber\\
&\times \Eext_j(\om,\bq+\bGp)
\end{align}
where again $\bq\in1\text{BZ}$. 

In considering dark matter-electron scattering we are primarily interested in the \emph{longitudinal} response. The longitudinal field is defined as $E_L(\omega,\bk) \equiv \bk\cdot \bE(\omega,\bk)/\left|\bk\right|$, where $\bk$ is a general momentum vector and not necessarily confined to the 1BZ. The transverse field is $\mathbf{E}_T\equiv \mathbf{E} - \left(\bk\cdot \bE(\omega,\bk)\right) \bk/\left|\bk\right|^2.$  The external field and dielectric tensor can be similarly decomposed. The scalar dielectric function is obtained by projecting both the total and external electric fields onto their longitudinal components: 
\begin{align}\label{eq:EL_total}
E_L(\om,\bq+\bG) \equiv &\sum_{\bGp} \epsilon^{-1}_{LL}(\omega,\bq+\bG,\bq+\bGp)\, \nonumber\\ &
\times \Eext_L(\om,\bq+\bGp). 
\end{align}
In \eqref{eq:EL_total} we have defined the (symmetrized\footnote{Note that there exists an alternative definition of the longitudinal dielectric function in the literature. If a factor of $\left|\bq+\bG\right|/\left|\bq+\bGp\right|$ is included on the right hand side of \eqref{eq:EL_total}, one arrives at the ``unsymmetrized longitudinal dielectric function''. The $\bG=\bGp$ elements of the symmetrized and unsymmetrized quantities are the same, but the off-diagonal components are not, so one should be careful to use a consistent definition in considering local field effects.}) longitudinal dielectric function 
\begin{align}\label{eq:eps_def}
 \epsilon^{-1}_{LL}(\omega,\bq+\bG,\bq+\bGp) \equiv& \frac{\left(q_i + G_i\right)\left(q_j+G^\prime_j \right)}{\left|\bq+\bG\right|\left|\bq+\bG'\right|}\nonumber\\
 &\times\epsilon^{-1}_{ij}(\omega,\bq+\bG,\bq+\bGp)
\end{align}
which describes the longitudinal response to a longitudinal external field. One can also define the matrices $\epsilon^{-1}_{LT, \, TL,\, TT}$ to describe the other components of the response, but for nearly isotropic crystals and the energies of interest for dark matter scattering, the purely longitudinal contribution dominates both $\mathbf{E}^{\rm ext}$ and the response. For compactness, we denote $\epsilon^{-1}_{\bG\bGp}(\omega,\bq)\equiv \epsilon^{-1}_{LL}(\omega,\bq+\bG,\bq+\bGp)$.

The microscopic dielectric function can be computed from the density response function $\chi_{\bfG \bfG'}(\bfq)$ in density functional theory, as described in Sec.~\ref{sec:DFT}. One can show (see e.g.~\cite{thesis}) that the susceptibility in the full interacting system is related to the KS susceptibility $\chi^{KS}$ via a Dyson equation
\begin{equation}\label{eq:chiDyson1}
\begin{aligned}
\chi_{\mathbf{G}\mathbf{G}'}&(\mathbf{q},\omega) = \chi^{KS}_{\mathbf{G}\mathbf{G}'}(\mathbf{q},\omega) + \sum_{\mathbf{G}_1,\mathbf{G}_2} \chi^{KS}_{\mathbf{G}\mathbf{G}_1}(\mathbf{q},\omega)  \\
&\times\left(\frac{4\pi \alpha_{em}}{\left| \mathbf{q}+\mathbf{G}_1 \right| \left|\mathbf{q}+\mathbf{G}_2 \right|}+f^{xc}_{\mathbf{G}_1\mathbf{G}_2}(\mathbf{q},\omega)\right)\\
&\times\chi_{\mathbf{G}_2\mathbf{G}'}(\mathbf{q},\omega)
\end{aligned}
\end{equation}
where $f^{xc}$ is a so-called ``exchange correlation kernel'' which is defined such that the charge density of the KS system exactly matches that of the full system. Exact knowledge of $f^{xc}$ would require solving for the wavefunctions of the full interacting system, however in TDDFT calculations one typically approximates this term using simple physically-motivated models such as the ``adiabatic local density approximation'' (ALDA), or dropping it altogether. Setting $f^{xc} \to 0$ in the expression above corresponds to the ``random phase approximation'' (RPA) which we use throughout this study. 

Furthermore, the susceptibility and polarizability are related by a separate Dyson equation
\begin{equation}\label{eq:chiDyson2}
\begin{aligned}
\chi_{\mathbf{G}\mathbf{G}'}&(\mathbf{q},\omega) = P_{\mathbf{G}\mathbf{G}'}(\mathbf{q},\omega) + \sum_{\mathbf{G}_1,\mathbf{G}_2} P_{\mathbf{G}\mathbf{G}_1}(\mathbf{q},\omega)  \\
&\times\left(\frac{4\pi \alpha_{em}}{\left| \mathbf{q}+\mathbf{G}_1 \right| \left|\mathbf{q}+\mathbf{G}_2 \right|}\right)\chi_{\mathbf{G}_2\mathbf{G}'}(\mathbf{q},\omega).
\end{aligned}
\end{equation}
Comparing \eqref{eq:chiDyson2} to \eqref{eq:chiDyson1} with $f^{xc} \to 0$, we see that in the RPA the polarizability of the full system is governed by the same Dyson equation as the KS susceptibility, motivating the approximation $P_{\bfG,\bfG'}(\bfq,\omega) \approx \chi^{KS}_{\bfG,\bfG'}(\bfq,\omega)$. This gives the RPA dielectric function of Eq.~\eqref{eq:Lindard_general} (see below), and reduces to that of Eq.~\eqref{eq:eps_DFT_appx} if the off-diagonal components are neglected and one restricts the momentum transfer to lie within the 1BZ.

\section{Comparison with previous works\label{app:formalismcomparison}}

The dynamic structure factor can be directly related to the DM scattering form factors appearing elsewhere in the literature when the Lindhard dielectric function (or random phase approximation) is used. In this section, we provide some additional formulae to help translate the presentation here in terms of $\epsilon(\omega,\bfk)$ to the results appearing in several previous studies. We also provide some plots comparing our results with those in Essig~et.~al.~\cite{Essig:2015cda} and Griffin~et.~al.~\cite{Griffin:2019mvc}.

\begin{figure*}[t]
\includegraphics[width=0.95\textwidth]{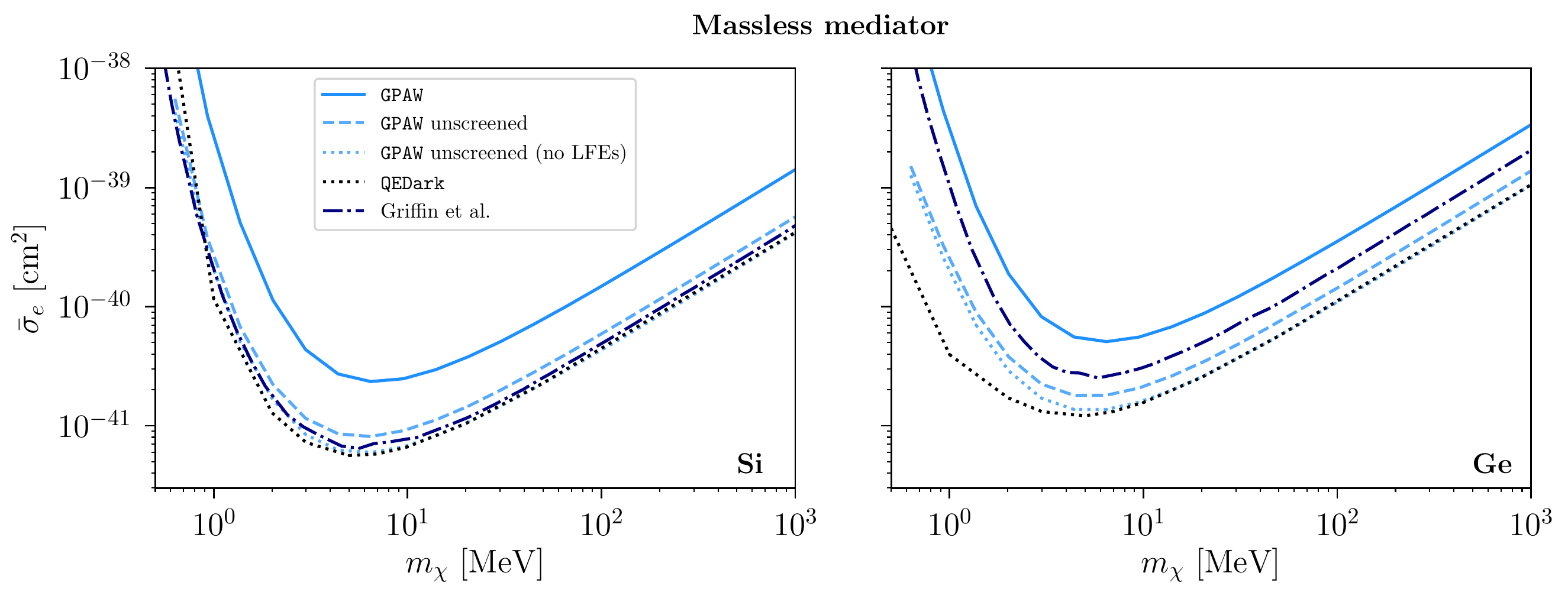}
\includegraphics[width=0.95\textwidth]{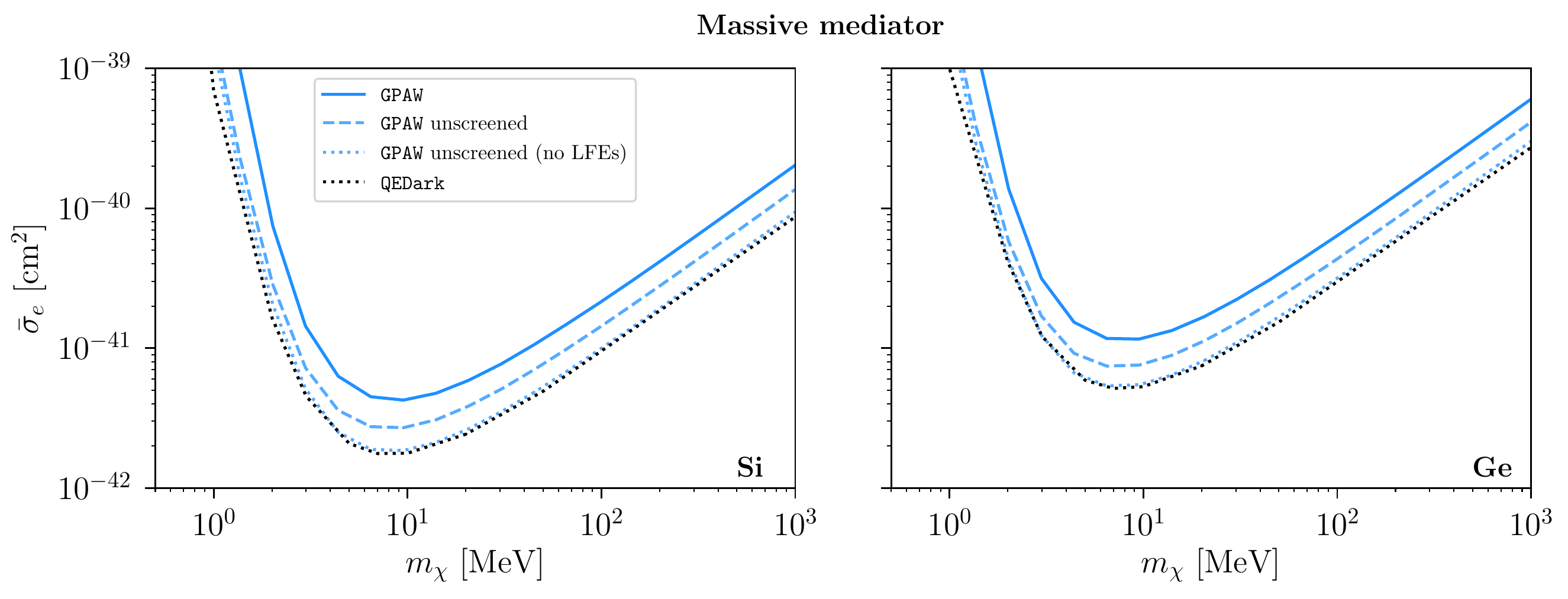}
\caption{ \label{fig:reachcompare} Comparison of the scattering reach with different calculations, assuming a $1e^-$ threshold with negligible background and kg-year exposure. The solid and dashed blue lines correspond to screened and unscreened rate obtained with our calculations of the dielectric function using \texttt{GPAW}, including local field effects. The dotted blue line shows the result if we compute the unscreened rate with the RPA dielectric function from \texttt{GPAW} without local field effects. This case is the one that corresponds closely to previous calculations, and numerically we find very good agreement with Griffin et al.~\cite{Griffin:2019mvc} and with \texttt{QEDark}~\cite{Essig:2015cda}.}
\end{figure*}

In the main text, the Lindhard dielectric function given in \eqref{eq:lindhard} was only valid for $\bfk$ within the 1BZ for a crystal. Here we generalize to account for reciprocal lattice vectors and split $\bfk = \bfq + \bfG$ where $\bfq$ lies in the 1BZ. The Lindhard dielectric function can be written as~\cite{Adler}
\begin{equation}
\begin{aligned}
    \epsilon_{\bfG \bfG'}^{\rm RPA}&(\bfq) =  \, \delta_{\bfG \bfG'} - \frac{4 \pi \alpha_{em}}{V} \frac{2}{ |\bfq + \bfG| |\bfq + \bfG'|} \times   
 \\
    	&  \lim_{\eta \to 0} \sum_{\bfp,\bfp',\ell, \ell'}  \frac{ f^{0}(\omega_{\bfp', \ell'}) - f^{0}(\omega_{\bfp,\ell}) }{ \omega_{\bfp', \ell'} - \omega_{\bfp, \ell} - \omega - i \eta } \, \eta_\bfG^* \eta_{\bfG'} \, \delta_{\bfp',\bfp+ \bfq},     
    \label{eq:Lindard_general}
\end{aligned}
\end{equation}
where the matrix elements above are defined by
\begin{align}
     \eta_\bfG \equiv \frac{1}{\Omega} \int_{\rm unit} d^3  \bfr \, u^*_{\bfp', \ell'}(\bfr) u_{\bfp, \ell}(\bfr) e^{i \bfG \cdot \bfr} 
\end{align}
with $\bfp$ within the 1BZ and $\bfG, \bfG'$ reciprocal lattice vectors. Here we have assumed that the real-space Bloch wavefunction for an electron in band $\ell$ can be written as
\begin{align}
	| \bfp, \ell \rangle  = \frac{e^{i \bfp \cdot \bfr}}{\sqrt{V}}  u_{\bfp, \ell} (\bfr) =  \sum_{\bfG} \frac{e^{i (\bfp + \bfG)\cdot \bfr}}{\sqrt{V}} u_\ell(\bfp + \bfG)
\end{align}
where $u_{\bfp, \ell}(\bfr)$ is periodic under under $\bfr \to \bfr + {\bf R}$, with ${\bf R}$ a lattice vector. In the second equality above, we have written the wavefunction in terms of the momentum-space coefficients $u_\ell(\bfp + \bfG)$. The matrix element above can equivalently be written in momentum space as
\begin{align}
     \eta_\bfG &= \sum_{\bfG_\Delta} u^*_{\ell'}(\bfp' + \bfG + \bfG_\Delta) u_\ell(\bfp + \bfG_\Delta) \\
      &\equiv f_{[\ell \bfp, \ell' \bfp',\bfG]}
\end{align}
where in the last line we make contact with the notation of Refs.~\cite{Essig:2015cda,Griffin:2019mvc}.

To compute the ELF, the dielectric matrix must be treated as a matrix in reciprocal lattice vectors and inverted to obtain the inverse dielectric function. In order to compare with results in the literature, we work in the approximation that the off-diagonal elements can be neglected, and restrict to $\bfG = \bfG'$ in \eqref{eq:Lindard_general}. Then taking the imaginary part of the dielectric function above gives
\begin{align}
    \Im ( &\epsilon^{\rm RPA}_{\bfG \bfG}(\bfq)) = \frac{4 \pi^2 \alpha_{em}}{V k^2} \sum_{\bfp,\bfp',\ell,\ell'}( f^{0}(\omega_{\bfp,\ell})  - f^{0}(\omega_{\bfp', \ell'}) ) \nonumber \\
    &\times \delta_{\bfp',\bfp+ \bfq} \, |f_{[\ell \bfp, \ell' \bfp',\bfG]}|^2 \, \delta(\omega_{\bfp', \ell'} - \omega_{\bfp, \ell} - \omega)
\end{align}
We use \eqref{eq:ELF_structurefactor}, take the continuum limit, and now explicitly include a factor of 2 for the spin sum. (This was implicit in equations in the main text.) We find that the structure factor can be written as
\begin{align}
    S(\omega,&\bfk) = \frac{2 \pi}{|\epsilon^{\rm RPA}(\omega,\bfk)|^2}  \sum_{\ell,\ell'} \int \! \! \frac{2\, d^3 \bfp}{(2\pi)^3}\frac{d^3 \bfp'}{(2\pi)^3}|f_{[\ell \bfp, \ell' \bfp',\bfG]}|^2 \,
     \nonumber \\
     & \times \sum_{\bfG} (2\pi)^3 \delta( \bfp +\bfk - \bfG - \bfp')  \delta(\omega_{\bfp', \ell'} - \omega_{\bfp, \ell} - \omega)  \nonumber  \\
    & \times f^{0}(\omega_{\bfp,\ell})(1 - f^{0}(\omega_{\bfp', \ell'}) )
\end{align}
where we introduce the sum over $\bfG$ to select out the piece of the incident DM momentum $\bfk$ that brings it to the first BZ. As noted before, this agrees with the definition of the structure factor in Ref.~\cite{Griffin:2019mvc} except for the $1/|\epsilon(\omega,\bfk)|^2$ screening factor appearing here.

To connect with the definitions in Ref.~\cite{Essig:2015cda}, which also averages over all directions in calculating the rate, we replace the 3-dimensional momentum delta function with a delta function averaged over the sphere, $\delta^{3}(\bfp - \bfp' 
+\bfk - \bfG) \to \delta(k - |\bfp' - \bfp + \bfG| )/(4 \pi k^{2})$. From this, we can immediately compare with the definition of the isotropic crystal form factor appearing there, and obtain \eqref{eq:fcrystal_vs_Imeps} by neglecting the factor of $(1 - e^{-\beta \omega})$ in the low temperature limit.

In Fig.~\ref{fig:reachcompare} we show a comparison of various calculations of the cross section reach, taking here $v_{\rm esc} = 600$ km/s, $v_0 = 230$ km/s, and $v_e = 240$ km/s for direct comparison. We show the screened and unscreened reach using our default calculations of the dielectric function, which account for local field effects, as discussed in Sec.~\ref{sec:DFT}. There is a small difference in the rate if we use the RPA dielectric function without local field effects (dotted light blue). This unscreened rate corresponds to the calculation of Refs.~\cite{Essig:2015cda,Griffin:2019mvc}, and with which our results agree very well. For Ge and scattering via massless mediators, there are somewhat larger differences in the unscreened rate across different calculations, which may be due to differences in the various DFT calculations (choice of exchange-correlation functionals, lattice constants, etc).

\begin{figure*}[t]
\includegraphics[width=0.95\textwidth]{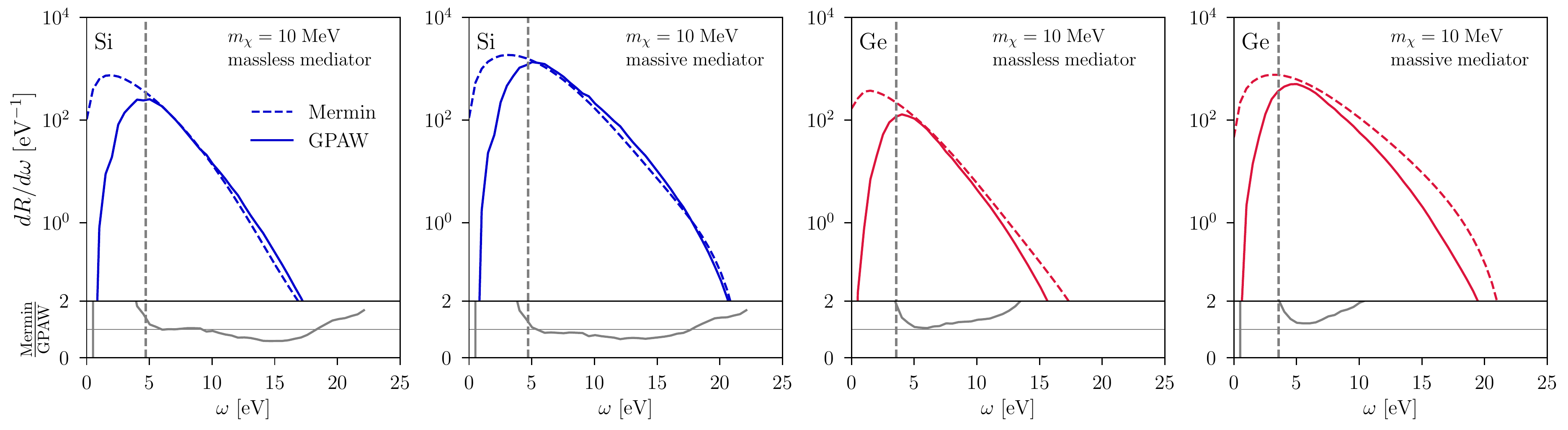}\\\vspace{0.15cm}
\includegraphics[width=0.95\textwidth]{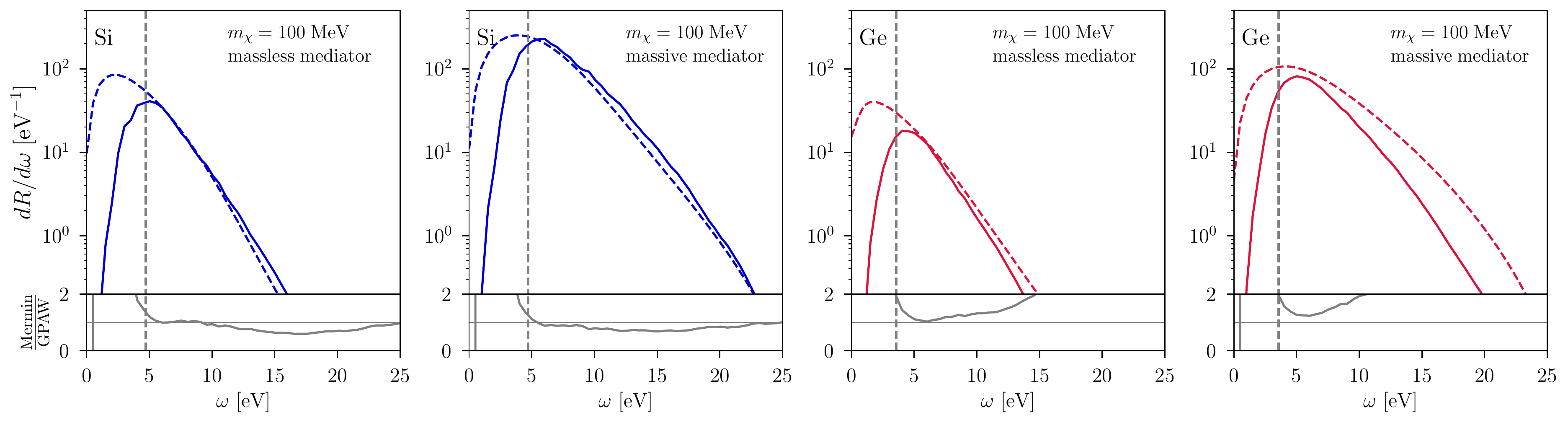}
\caption{Comparison of the differential scattering rate as obtained with a DFT calculation (GPAW) and with the Mermin oscillator method (Mermin), for $\bar \sigma_e=10^{-38}\;\text{cm}^2$ and a kg-year exposure. The vertical dashed line indicates the 2$e^-$ threshold.  \label{fig:mermindifferential} }
\end{figure*}

\begin{figure*}[t]
\includegraphics[width=0.95\textwidth]{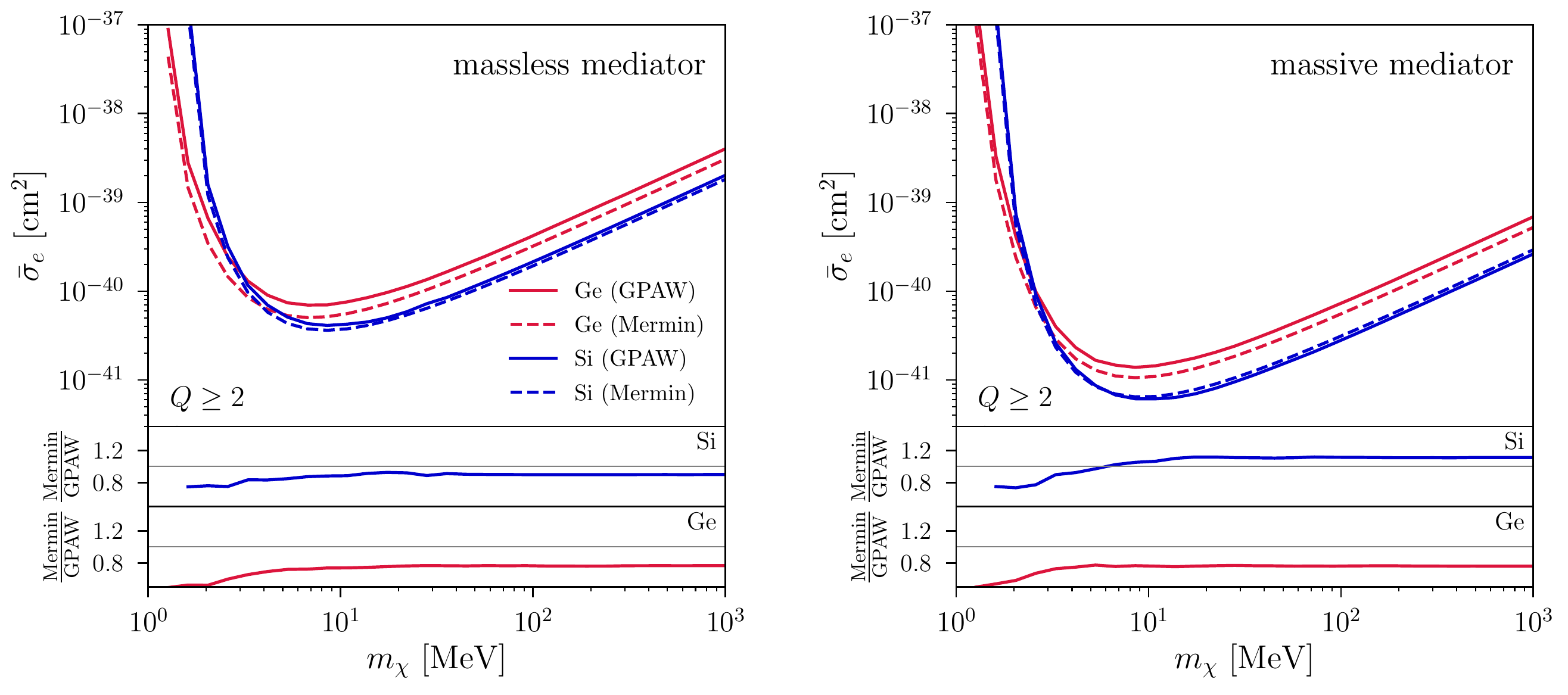}
\caption{Comparison of the reach between the calculations using density functional theory (\texttt{GPAW}) and the Mermin oscillator method, assuming a $2e^{-}$ threshold with negligible background and 1 kg-year exposure. 
\label{fig:method_comparison}}
\end{figure*}

\section{Mermin oscillator results\label{sec:comparison}}
In this appendix we briefly present some results obtained with the Mermin oscillator method (see Sec.~\ref{sec:mermin}), and study how they compare with those obtained with the DFT calculation (see Sec.~\ref{sec:DFT}).  With both methodologies, accessing the high-$k$ regime is challenging, for different reasons. In the DFT calculation, an increasingly large basis set of wave functions is needed, which increases the computational cost of the calculation. In the Mermin oscillator method, the high $k$ regime corresponds to a substantial extrapolation from the experimental data, which was taken in the optical limit ($k=0$).  For $k\gtrsim12$ keV our numerical results with the Mermin oscillator method in particular cease to be stable and we therefore impose a cut of $k< 12$ keV on the phase space in both calculations. We verified that the contribution of the omitted part of the phase space is negligible in the integrated rate, but it slightly affects the shape of $dR/d\omega$ for $\omega\gtrsim 15$ eV..

With this assumption, Fig.~\ref{fig:mermindifferential} shows the differential scattering rate obtained with both methods. We find overall good agreement, except for low and high $\omega$. Poor agreement at low $\omega$ is anticipated, since the Mermin oscillator method models the semiconductor as a linear combination of free electron gas systems, and is therefore expected to be less reliable for $\omega$ near the band gap of the material.  Once we impose the $2e^-$ threshold (dashed line), the agreement between both methods is largely satisfactory. The substantial deviations in the high $\omega$ regime are also straightforward to understand. For kinematical reasons, this regime corresponds to the higher $k$ part of the phase space, which is challenging for both methods as discussed above. Further studies are needed to bring down the uncertainty in this region. On the other hand, for experiments with a 2$e^-$ threshold, this region provides a subdominant contribution to the rate, and is likely only relevant in the event of a discovery.

The integrated rate above the $2e^-$ threshold is shown in Fig.~\ref{fig:method_comparison}. For Si, both methods agree to within 10\%, and for Ge the agreement is within roughly 30\% in most of the mass range. The uncertainties increase for low masses, due to the challenge of modeling the ELF accurately for $\omega$ close to the band gap.

\bibliography{dielectric}

\end{document}